\documentclass[%
 aps,
 pre,
 amsmath,amssymb,
 reprint,%
nofootinbib
]{revtex4-1}

\usepackage{graphicx}% Include figure files
\usepackage{dcolumn}% Align table columns on decimal point
\usepackage{bm}% bold math
\usepackage[utf8]{inputenc}
\usepackage[T1]{fontenc}
\usepackage{siunitx}
\usepackage{tikz}
\usetikzlibrary{plotmarks}
\usepackage{subcaption}
\usepackage{xcolor}

\begin{document}

\title{From hydrodynamics to dipolar colloids: modeling complex interactions and self-organization with generalized potentials}
\author{T.J.J.M. van Overveld}
\affiliation{Fluids and Flows group and J.M. Burgers Center for Fluid Mechanics, Department of Applied Physics and Science Education, Eindhoven University of Technology, P.O. Box 513, 5600 MB Eindhoven, The Netherlands}
% \affiliation{Transport Phenomena group, Department of Chemical Engineering, Delft University of Technology, Delft, The Netherlands}
\author{W.G. Ellenbroek}
\affiliation{Soft Matter and Biological Physics group, Department of Applied Physics and Science Education, Eindhoven University of Technology, P.O. Box 513, 5600 MB Eindhoven, The Netherlands}
\author{J.M. Meijer}
\affiliation{Soft Matter and Biological Physics group, Department of Applied Physics and Science Education, Eindhoven University of Technology, P.O. Box 513, 5600 MB Eindhoven, The Netherlands}
\author{H.J.H. Clercx}
\affiliation{Fluids and Flows group and J.M. Burgers Center for Fluid Mechanics, Department of Applied Physics and Science Education, Eindhoven University of Technology, P.O. Box 513, 5600 MB Eindhoven, The Netherlands }
\author{M. Duran-Matute}
\email{m.duran.matute@tue.nl}
\affiliation{Fluids and Flows group and J.M. Burgers Center for Fluid Mechanics, Department of Applied Physics and Science Education, Eindhoven University of Technology, P.O. Box 513, 5600 MB Eindhoven, The Netherlands }

\date{\today}

\begin{abstract}
    The self-organization of clusters of particles is a fundamental phenomenon across various physical systems, including hydrodynamic and colloidal systems. 
    One example is that of dense spherical particles submerged in a viscous fluid and subjected to horizontal oscillations. The interaction of the particles with the oscillating flow leads to the formation of one-particle-thick chains or multiple-particle-wide bands, both oriented perpendicular to the oscillation direction.
    In this study, we model the hydrodynamic interactions between such particles and parallel chains using simplified potentials. We first focus on the hydrodynamic interactions between chains, which we characterize using data from fully resolved numerical simulations. Based on these interactions, we propose a simplified model potential, called the \emph{Siren} potential, which combines the representative hydrodynamic interactions: short-range attraction, mid-range repulsion, and long-range attraction.
    Through one-dimensional Monte Carlo simulations, we successfully replicate the characteristic patterns observed in hydrodynamic experiments and draw the phase diagram for the model potential. We further extend our analysis to two-dimensional systems, introducing a \emph{dipole-capillary} model potential that accounts for both chain formation and \emph{Siren}-like chain interactions. This potential is based on a system with colloidal particles at an interface, where chain formation is driven by an external electric field that induces a dipole moment parallel to the interface in each particle. The capillary force contributes the long-range attraction.
    Starting with parallel chains, the patterns in the two-dimensional Monte Carlo simulations of this colloidal system are similar to those observed in the hydrodynamic experiments. However, we identify that nonlinear interactions are important for some distinct steps in the chain formation. 
    Still, the model potentials help clarify the dynamic behavior of the particles and chains due to the complex interactions encountered in both hydrodynamic and colloidal systems, drawing parallels between them. 
\end{abstract}

\maketitle

\section{Introduction}\label{sec:introduction}
    The self-organization of clusters of particles is a common and important feature in a wide range of systems across multiple orders of magnitude in space and time. A few notable examples include the formation of ripples in sediment bedforms \citep{kennedy1969formation}, the collective dynamics of microswimmers \citep{bechinger2016active}, and flows of granular materials \citep{silbert2002boundary}.
    Recently, there has been a renewed interest in exploring the connections between such different systems in an interdisciplinary approach \citep{araujo2023steering}. The aim is to enhance the understanding of self-organization phenomena and the influence of confinement by drawing parallels across different scientific disciplines.
    
    A particular case of self-organization occurs in electrorheological fluids, where colloidal particles suspended in a liquid medium are subjected to an external oscillating electric field. Such systems have been studied extensively a few decades ago (see, e.g., the reviews by \citep{gast1989electrorheological,halsey1992electrorheological}).
    The electric field induces a dipole moment in each particle, causing them to interact and form chains within a few milliseconds \citep{gast1989electrorheological,halsey1992electrorheological,yethiraj2002monodisperse}. These chains align with the external field and interact with other chains. They repel each other at large distances but also attract at very short distances, forming stable clusters \citep{almudallal2011simulation}. Such chain formation also occurs for particles with intrinsic dipole moments \citep{weis1993chain}. However, in such a case, the chains do not align along a preferential direction due to the lack of external forcing and its associated directionality. Similar self-organization phenomena are also observed in other colloidal systems with charged colloids or colloid–polymer mixtures \citep{royall2018hunting}.
    
    The interactions in these colloidal systems are often modeled using SALR (short-range attractive and long-range repulsive) potentials \citep{liu2019colloidal}. Such potentials are sometimes referred to as \emph{Mermaid} potentials due to their attractive head and repulsive tail (as beautifully demonstrated by \citet{hooshanginejad_mermaid_2023}). 
    Typical examples of SALR models are the hard-core double-Yukawa (HCDY) potential, used for studying equations of state in hard-core fluids \citep{lin2002monte}, and the combined Lennard-Jones and Yukawa potentials, used for studying clustering phenomena in fluids and gels \citep{sciortino2004equilibrium}.
    The combination of short-range attraction and long-range repulsion in systems governed by SALR interactions can lead to clustering and phase separation \citep{liu2019colloidal,munao2022competition}. For clustering, particles must overcome a potential barrier, possibly due to (thermal) fluctuations or driven by confinement \citep{araujo2023steering}.

\begin{figure*}
        \includegraphics[width=\textwidth]{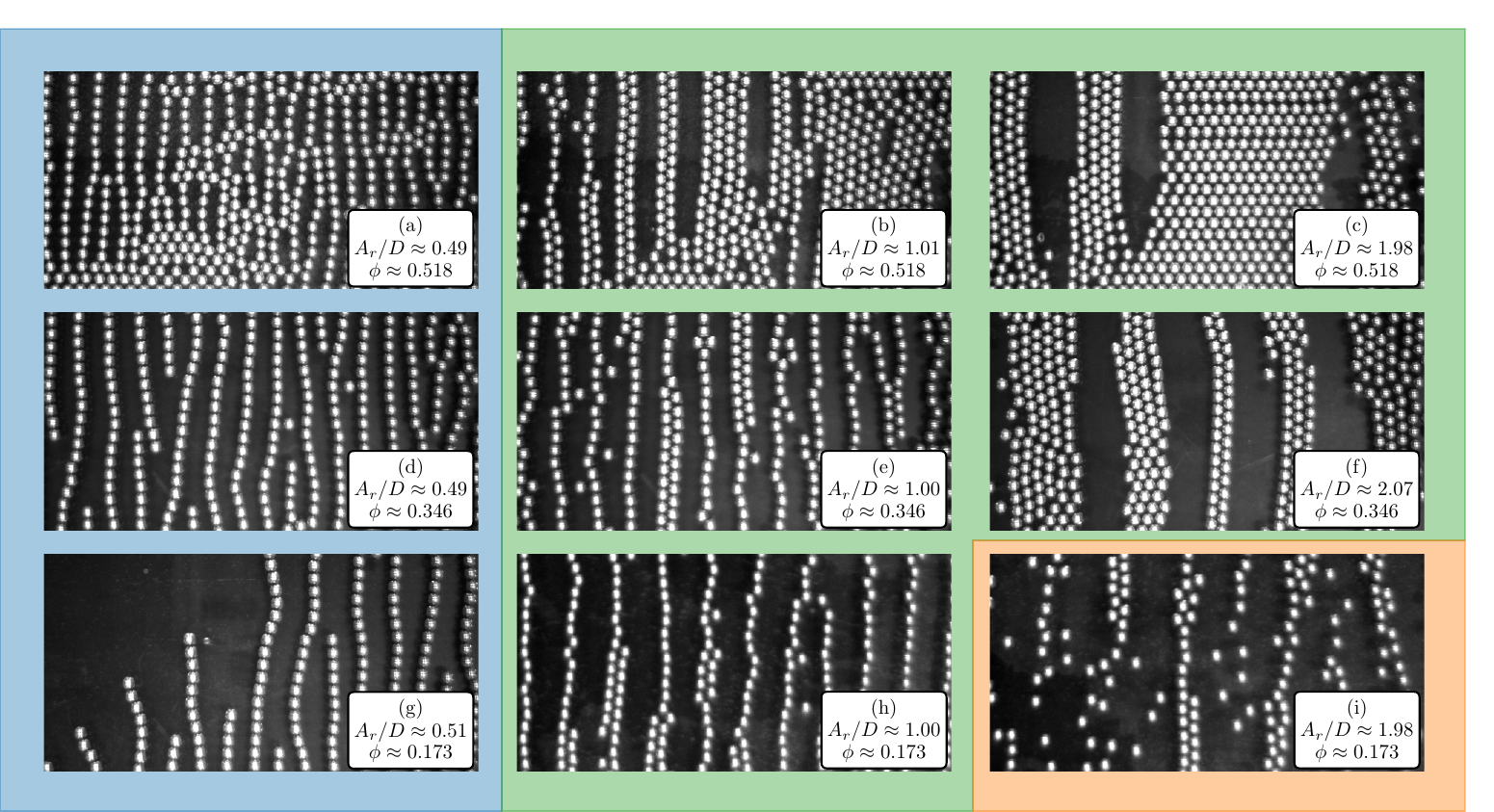}
        \caption{Examples of the patterns observed in the experiments by \citet{overveld2023pattern}, as a function of the relative particle-fluid excursion length normalized by particle diameter, $A_r/D$, and particle coverage fraction $\phi$. The patterns range from one-particle-thick chains to multiple-particle-wide bands, all with an intrinsic spacing between them that varies with $A_r/D$. The colors indicate different regions in the parameter space, with only chains and irregular clusters (blue), chains and bands (green), and more disordered structures (orange).}
        \label{fig:example-of-patterns}
    \end{figure*}

    Self-organization can also be found in hydrodynamic systems, driven by the particle-fluid interactions and nonlinearities in the flow \citep{wunenburger2002periodic,voth2002ordered,klotsa2009chain}. A good example are the experiments by \citet{overveld2023pattern}, where patterns are formed by spherical particles inside an oscillating box filled with viscous fluid and observed a wide range of distinct patterns. These patterns range from one-particle-thick chains to multiple-particle-wide bands, all with an intrinsic spacing between them that varies with the oscillatory forcing. Note that this hydrodynamic system is macroscopic, such that thermal fluctuations do not play a role in the particle behavior. An overview of these patterns is given in Fig.~\ref{fig:example-of-patterns}, representing the parameter space as a function of the relative particle-fluid excursion length (i.e., how far the particle moves with respect to the ambient fluid) $A_r$ normalized by the particle diameter $D$ and the particle coverage fraction $\phi$.
        
    Complementary numerical simulations revealed that the hydrodynamic interactions are due to vortices in the period-averaged flow. These vortices vary in size, position, and magnitude with different flow conditions, adding complexity to the interactions. Overall, the interactions can be divided into a short-range attraction, a mid-range repulsion, and a long-range attraction. The first two interactions are consistent with the SALR model. However, the long-range attraction is an additional effect, leading to compact patterns that do not necessarily spread out over the entire domain. The balance between mid-range repulsion and long-range attraction determines the intrinsic spacing between the chains or bands. 
    Notably, the long-range attraction emerges due to the hydrodynamic interaction between the particle chains and the surrounding fluid, becoming apparent only when the chains are formed. Therefore, it can be described as `self-reinforced confinement' since it arises from the collective dynamics and internal feedback loops within the system itself \citep{araujo2023steering}. 
    
    By combining the experiments and simulation data, three distinct regions were identified, as illustrated in Fig.~\ref{fig:example-of-patterns}. For $A_r/D \lesssim 0.7$, one-particle-thick chains form, accompanied by some irregular clusters if $\phi$ is relatively large. For $A_r/D\approx2$ and small $\phi$, many isolated particles and only a few large structures, primarily two-particle-wide bands, were found. The patterns in the rest of the parameter space consist of combinations of chains and bands, characterized by an increasing distance between them with increasing $A_r/D$ and a typical width that increases with both $A_r/D$ and $\phi$ (see \citet{overveld2023pattern} for a detailed discussion).
    
    In this study, we capture the essential aspects of the complex self-organization of spherical particles in an oscillating flow, using a simplified model potential, that we name the \emph{Siren} potential\footnote{Like the Mermaid potential, the name \emph{Siren potential} is inspired by (Greek) mythology. Although \emph{mermaid} and \emph{siren} are now sometimes used as synonyms, sirens were first described as mythological beings with attractive human heads and repulsive birdlike bodies. They lure their prey from far away using their voices \citep{pollard1952muses}.}, which incorporates four distinct interactions: hard-sphere repulsion, short-range attraction, mid-range repulsion, and long-range attraction.
        
    By independently tuning the relative interaction strengths in the Siren potential, we replicate self-organization phenomena observed in hydrodynamic or colloidal systems. This potential allows us to explore the full parameter space without constraints. 
    This level of control of the individual interactions is typically not feasible in physical systems, since interactions are coupled to each other and depend on the external forcing in a complex manner. For example, in colloidal systems, the dipole interactions that induce short-range attraction, leading to clustering, also cause long-range repulsion. In the hydrodynamic system, the vortices in the period-averaged flow underlie the chain and band formation (through short-range attraction) and their mutual interactions (through mid-range repulsion and long-range attraction). The individual interactions can, therefore, not be tuned individually.
    
    The Siren potential is a generalization encompassing other commonly observed potentials in various systems. Through different parameter combinations, we can qualitatively reproduce potentials resembling a SALR (Mermaid) potential, a potential well (e.g., similar to Lennard-Jones), and a purely attractive potential. 
    Furthermore, the Siren potential shares similarities with potentials used in Derjaguin-Landau-Verwey-Overbeek (DLVO) theory, which describes interactions between charged colloidal particles in aqueous dispersions \citep{verwey1948theory}. 
    In DLVO theory, electrostatic repulsion (dominant at mid-range) and Van der Waals attraction (dominant at short and long-range) contribute to the interaction potential. The combination results in a deep primary minimum at a short range and a shallow secondary minimum at a larger range, corresponding to irreversible coagulation and reversible flocculation, respectively \citep{salager1994interfacial}.

    A systematic approach is followed to reach our aim of capturing the essential aspects of the complex self-organization processes. We first apply the Siren potential to a one-dimensional model system, effectively describing the interactions between parallel chains in the experiments conducted by \citet{overveld2023pattern}. Even within this simplified framework, we show that a rich range of patterns can be found. The equilibrium states are predicted using analytical calculations based on lattice sums. We further explore the phase space and the pattern's sensitivity to initial conditions using one-dimensional Monte Carlo simulations. 
    Then, we extend to two-dimensional cases to also consider the interactions between individual particles and the formation of the chains. We propose a novel approach for realizing the Siren potential in two dimensions based on common interactions in colloidal systems with dipolar particles (in an external electric field) and an interface between two fluids (exploiting capillary forces). We explore the self-organization in such systems using two-dimensional Monte Carlo simulations, including the effects of initial conditions and confinement. We highlight similarities and differences between this colloidal system and the hydrodynamic system by \citet{overveld2023pattern}.
    
    We explore the interactions between parallel chains in the hydrodynamic system in Sec.~\ref{sec:hydrodynamics}. Then, in Sec.~\ref{sec:modelpotential}, we introduce the Siren potential along with an analysis of the equilibrium states. The Monte Carlo results for one-dimensional systems are shown in Sec.~\ref{sec:MC1D}, and the connection to colloidal systems is explored in Sec.~\ref{sec:MC2D}. Finally, the conclusions are drawn in Sec.~\ref{sec:conclusion}.

\section{Interactions in the hydrodynamic system}\label{sec:hydrodynamics}
    As a starting point, we consider numerical simulations of two particle chains in an oscillating flow, following the approach by \citet{overveld2023pattern}. The two chains are oriented along the $y$-axis, which is perpendicular to the oscillation direction along the $x$-axis. The chains are separated by a distance $\lambda$. The particles, with diameter $D$, are kept fixed in space. The flow around each particle is fully resolved. The particle-fluid interactions are accounted for using an immersed boundary method, based on the code by \citet{breugem2012second}. This code was previously used in related studies involving spherical particles in oscillatory flows \cite{overveld2022numerical,overveld2022effect}.

    In our simulations, we place two particle chains (each consisting of ten particles) on a flat bottom within a doubly periodic domain of size $10D\times20D$ (along and perpendicular to the chains, respectively), filled with a Newtonian fluid with density $\rho_f$. The fluid oscillates with excursion amplitude $A_r$ and angular frequency $\omega$. The top and bottom boundaries, separated by a distance $H=5D$ (as in the experiments) and characterized by no-slip/no-penetration conditions, oscillate with the same amplitude and frequency as the fluid to simulate an oscillating box. 
    
\begin{figure}
        \includegraphics[width=\linewidth]{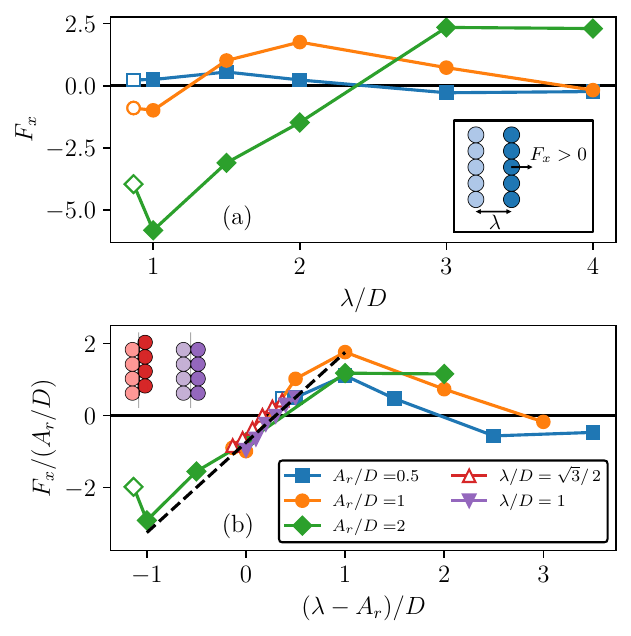}
        \caption{(a) The dimensionless, average streamwise force per particle $F_x$ as a function of the normalized chain spacing $\lambda/D$, where positive values of $F_x$ correspond to repulsive interactions (see insert). The chains are aligned (as in the insert), except for the cases with open symbols, which correspond to touching chains in a staggered configuration. (b) Collapse of the data displayed in (a). The triangles represent forces from simulations of two touching chains in staggered ($\lambda/D=\sqrt{3}/2$, upward triangle) or straight ($\lambda/D=1$, downward triangle) configuration (as shown in the upper left corner), for $A_r/D=[0.5,0.6,\dots,1]$. Around the first zero crossing, the data roughly follows the black dashed line, given by Eq.~\eqref{eq:linearfit}.}
        \label{fig:forces_chains}
    \end{figure}

    We calculate the period-averaged forces acting on the particles within one of the two parallel chains. Specifically, we consider the forces in the oscillation direction since these provide insight into the interaction between the chains and their equilibrium spacing. The average streamwise force on each particle is nondimensionalized by $\rho_f\left(\pi D^3/6\right)A_r\omega^2$ and denoted as $F_x$.
    Figure~\ref{fig:forces_chains}(a) shows $F_x$ on each particle as a function of the normalized chain spacing $\lambda/D$ and normalized oscillation amplitude $A_r/D$.
    Negative values of $F_x$ correspond to attraction between the chains, while positive values correspond to repulsion. We observe two important aspects in Fig.~\ref{fig:forces_chains}(a): first, the maximum values of the forces increase with $A_r/D$, and second, the maximum force shifts towards larger $\lambda/D$ values as $A_r/D$ increases. 
    
    Next, we rescale $F_x$ with $A_r/D$ and subtract $A_r/D$ from $\lambda/D$ so that the data roughly collapses on a single curve. The result is shown in Fig.~\ref{fig:forces_chains}(b). This rescaling approach aligns with previous findings on the period-averaged flow fields \cite{overveld2023pattern}, where the vortices grow stronger with increasing $A_r/D$, justifying the rescaling of $F_x$ with $A_r/D$. 
    Additionally, the size of these vortices is roughly proportional to $A_r/D$, such that their centers shift linearly away from the chains with increasing $A_r/D$.
    Since the vortex is the active interaction center, the normalized distance between particle and vortex, $(\lambda-A_r)/D$, determines the (relative) magnitude of the instantaneous interaction, justifying the rescaling of the horizontal axis.

    The rescaled data in Fig.~\ref{fig:forces_chains}(b) give a comprehensive overview of the interactions between two chains, revealing three distinct regimes. Upon increasing the value of $(\lambda-A_r)/D$, the interaction is first attractive, then repulsive, and finally, attractive again. Moreover, the data around the first zero crossing ($(\lambda-A_r)/D\approx0.3$) collapse onto a straight line which we approximate by
    \begin{equation}\label{eq:linearfit}
        \frac{F_x}{A_r/D} = -0.75+2.5\frac{\left(\lambda-A_r\right)}{D},
    \end{equation}
    shown as the black dashed line in Fig.~\ref{fig:forces_chains}(b). Based on Eq.~\eqref{eq:linearfit} and the collapsed data, we estimate that short-range attraction occurs when $\lambda/D\lesssim 0.3+A_r/D$. This means that two touching chains (i.e., with $\lambda/D=1$) form a stable two-particle-wide band when $A_r/D\gtrsim0.7$, which is in complete agreement with the results of \citet{overveld2023pattern}.
    
    Note that in Fig.~\ref{fig:forces_chains}(a), the leftmost data point for each value of $A_r/D$ (with an open symbol) does not follow the linear trend described by Eq.~\eqref{eq:linearfit}. These points correspond to two touching chains in a staggered configuration, where $\lambda/D=\sqrt{3}/2$. The magnitude of the forces in this staggered configuration is smaller than in the straight configuration, which is especially evident for large $A_r/D$, e.g., for the leftmost green data point. The discrepancy between the staggered and straight configurations also appears in the offset between the red and purple triangles in Fig.~\ref{fig:forces_chains}(b).

\section{Model potential for the hydrodynamic interactions between parallel chains}\label{sec:modelpotential}
    We now describe the interactions from the previous section in a more generalized manner. We use lattice sums to analyze the energy associated with specific configurations to gain insight into the equilibrium states of particles within a given potential. For the sake of simplicity, we solely focus on a one-dimensional pattern within an infinitely large domain. This particular case corresponds to the streamwise interactions between (non-staggered) chains in the hydrodynamic system. Essentially, each particle within the one-dimensional model system represents a chain in the hydrodynamic system.

    \subsection{Siren potential}\label{sec:sirenpotential}    
        We introduce an effective potential $U$ that qualitatively captures the essential aspects of the hydrodynamic interactions between parallel chains. The potential has three terms and is given by
        \begin{equation}\label{eq:sirenpotential}
            U(r/D) = -\frac{A}{\left(r/D\right)^2}+\frac{B}{\left(r/D\right)^4}-\frac{C}{\left(r/D\right)^6},
        \end{equation}
        where $r/D$ is the normalized distance between two particles in the oscillation direction, and $A$, $B$, and $C$ are positive constants, representing the magnitude of long-range attraction, mid-range repulsion, and short-range attraction, respectively. 
        
        This model is remarkably versatile since it can reproduce potentials that are similar to those in other studies. We can characterize different types of potential based on the relative magnitude of the terms, $A/B$ and $C/B$, as shown in Fig.~\ref{fig:ab_cb}.
        Details on the boundaries between the regions are given in the Appendix \ref{sec:appendix}.
        
        \begin{figure}
            \includegraphics[width=\linewidth]{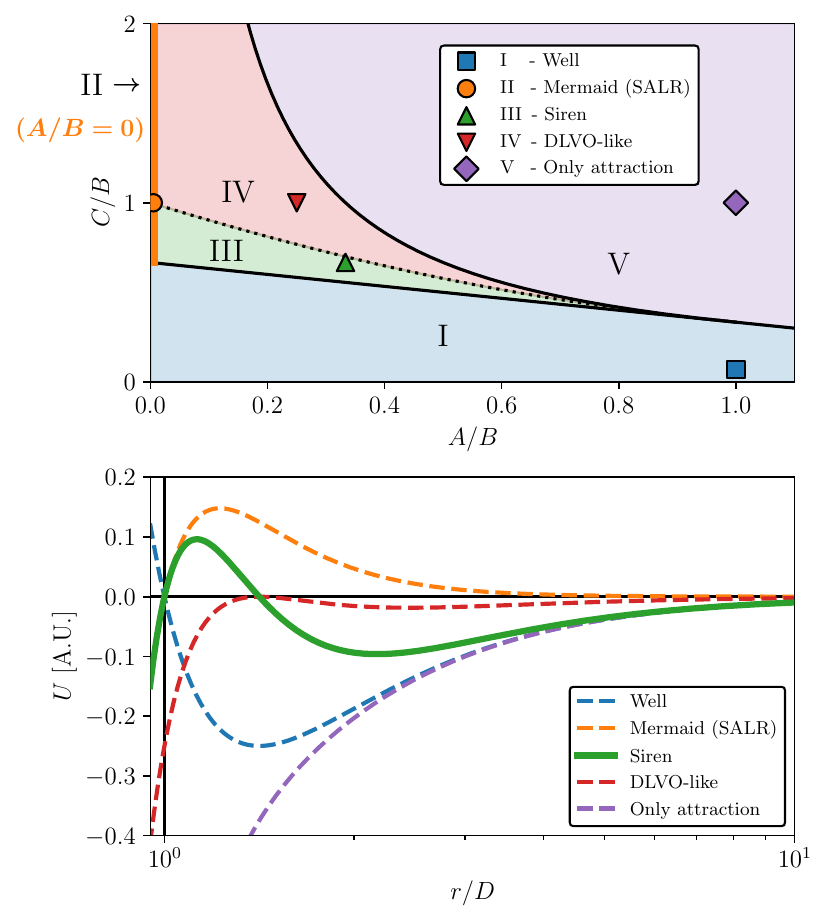}
            \caption{(a) The parameter space in terms of $A/B$ and $C/B$ that determines the characteristics of the potential in Eq.~\eqref{eq:sirenpotential}. A net short-range attraction only occurs for sufficiently large $C/B$ (above the lower black line). The interaction is attractive at every distance if the magnitudes of the two attractive terms are too large (above the upper black line). Above the dotted line, the potential is similar to those commonly used in DLVO theory, with the primary minimum (related to irreversible clustering) below the secondary minimum. (b) Examples of the potentials as indicated by the colored symbols in (a). The Siren potential (solid green curve), characterized by $[A,B,C]=[1,3,2]$, is used for the rest of this section. The vertical black line at $r/D=1$ represents the shortest possible center-to-center distance between two hard spheres.}
            \label{fig:ab_cb}
        \end{figure}
        
        In the subsequent analysis, we focus on the specific case with $[A,B,C]=[1,3,2]$, as shown in Fig.~\ref{fig:ab_cb}. For these specific values, the maximum is found at $\left(r_\mathrm{max}/D,U_\mathrm{max}\right)=\left(\sqrt{3-\sqrt{3}},1/(6\sqrt{3})\right)\approx(1.13,0.096)$ and the (secondary) minimum is found at ${(r_\mathrm{min}/D,U_\mathrm{min})=\left(\sqrt{3+\sqrt{3}},-1/(6\sqrt{3})\right)\approx(2.18,-0.096)}$. The potential is attractive for $r < r_\mathrm{max}$, repulsive for $r_\mathrm{max} < r < r_\mathrm{min}$, and attractive for $r>r_\mathrm{min}$.
        As shown in Fig.~\ref{fig:ab_cb}(b), the short-range behavior is similar to that of a \emph{Mermaid} potential (with $[A,B,C]=[0,1,1]$), while the long-range behavior is similar to a potential well (with $[A,B,C]=[1,1,0]$).

    \subsection{Lattice sums}\label{sec:latticesums}
        We consider lattice sums for more fundamental insight into the connection between the previously introduced potential and the (one-dimensional) equilibrium particle configurations. A more in-depth treatment of lattice sums is given by \citep{borwein2013lattice}. Note that the analysis in this section focuses on novel results rather than an overview of the literature.
        The total energy inside the system is given by the sum of all interactions between the particles,
        \begin{equation}\label{eq:totalenergy}
            \begin{split}
                \Bar{U} &= \frac{1}{2}\sum_{i=1}^N\sum_{\substack{j=1\\ j\neq i}}^N U(r_{ij}/D) \equiv N u, \\
            \end{split}
        \end{equation}
        where $N$ is the total number of particles in the system, $u$ is the average energy per particle, and the factor $1/2$ corrects for double counting of particle pairs.
    
        \begin{figure}
            \includegraphics[width=\linewidth]{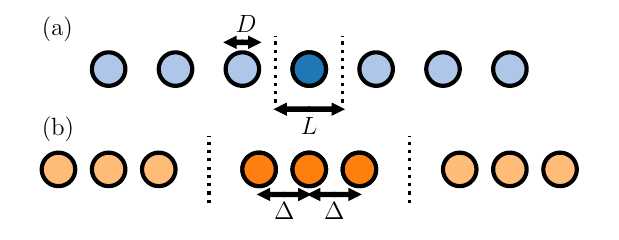}
            \caption{The lattice sums are calculated for two distinct one-dimensional configurations: (a) a uniform particle distribution and (b) groups of $M$ particles with an equal spacing $\Delta$. In the example configuration shown here, $M=3$. The particle diameter is $D$ and the unit cell length $L$.}
            \label{fig:schematic}
        \end{figure}
        
        We now apply this energy calculation to two distinct particle arrangements schematically shown in Fig.~\ref{fig:schematic}. We first consider the uniform particle distribution, shown in Fig.~\ref{fig:schematic}(a), since it is conceptually and mathematically straightforward. Then, we consider groups of particles with a specific interparticle spacing, shown in Fig.~\ref{fig:schematic}(b), which are more complex but also more versatile.
        For relatively large values of $M$, the configuration can be interpreted as a pattern of parallel chains with equilibrium spacing $\Delta$. Alternatively, when $\Delta = D$ (i.e., clusters of touching particles), the configuration resembles bands that are $M$ particles wide, with the spacing between these bands depending on $L$. Finally, note that for $M=1$, case (b) is equivalent to (a).

        \subsubsection{Uniform particle distribution}
            For a uniform distribution, we consider a particle with size $D$ in a unit cell of length $L$, as shown in Fig.~\ref{fig:schematic}(a). The energy per particle is given by
            \begin{widetext}
                \begin{equation}\label{eq:u1}
                    \begin{split}
                        u_0 &= \frac{1}{2}\sum_{\substack{j=-\infty\\j\neq 0}}^{\infty}U(jL/D),    \\
                        &= -\frac{A}{2(L/D)^2}\sum_{\substack{j=-\infty\\j\neq 0}}^{\infty}\frac{1}{j^2}+\frac{B}{2(L/D)^4}\sum_{\substack{j=-\infty\\j\neq 0}}^{\infty}\frac{1}{j^4}-\frac{C}{2(L/D)^6}\sum_{\substack{j=-\infty\\j\neq 0}}^{\infty}\frac{1}{j^6},\\
                        &= -A\frac{\pi^2\phi^2}{6}+ B\frac{\pi^4\phi^4}{90}- C\frac{\pi^6\phi^6}{945},\\
                    \end{split}
                \end{equation}
            \end{widetext}
            where $\phi=D/L$ is the portion of the unit cell occupied by the particle. The treatment of the infinite sums is described in, e.g., \citep{courant1999infinite}.
            The three terms on the bottom line in Eq.~\eqref{eq:u1} correspond to those in Eq.~\eqref{eq:sirenpotential}, demonstrating that the energy per particle, unsurprisingly, depends on the strengths of the separate interactions.
        
        \subsubsection{Groups of particles with a specific interparticle spacing}
            Next, we consider a group of $M$ particles with the centers between neighboring particles separated by a distance $\Delta$ within a unit cell of length $L$, as shown in Fig.~\ref{fig:schematic}(b). The average energy per particle is given by 
            \begin{widetext}
                \begin{equation}\label{eq:Mclustersdelta}
                    \begin{split}
                        u_M &= \frac{1}{2}\sum_{\substack{j=-\infty\\j\neq 0}}^{\infty} U(jL/D)+\frac{1}{M}\sum_{m=1}^{M-1} m\sum_{j=-\infty}^{\infty}U(jL/D+(\Delta/D)(M-m)) \\ 
                        &=  u_0 +\frac{1}{M}\sum_{m=1}^{M-1}\biggl[-\frac{A}{(L/D)^2}\Psi\left(1,\frac{\Delta}{L}(M-m)\right)+\frac{B}{6(L/D)^4}\Psi\left(3,\frac{\Delta}{L}(M-m)\right) -\frac{C}{120(L/D)^6}\Psi\left(5,\frac{\Delta}{L}(M-m)\right)\biggr],
                    \end{split}
                \end{equation}
            \end{widetext}
            where
            \begin{equation}
                \Psi(n,x) \equiv \psi(n,x)+\psi(n,1-x),
            \end{equation}
            with $\psi(n,x)$ the polygamma function of order $n$, defined as
            \begin{equation}
                \psi(n,x)\equiv \frac{d^{n+1}}{dx^{n+1}}\ln\Gamma(x),
            \end{equation}
            with $\Gamma(x)$ the Gamma function \citep{NIST:DLMF}.
            
\begin{figure*}
                \includegraphics[width=\linewidth]{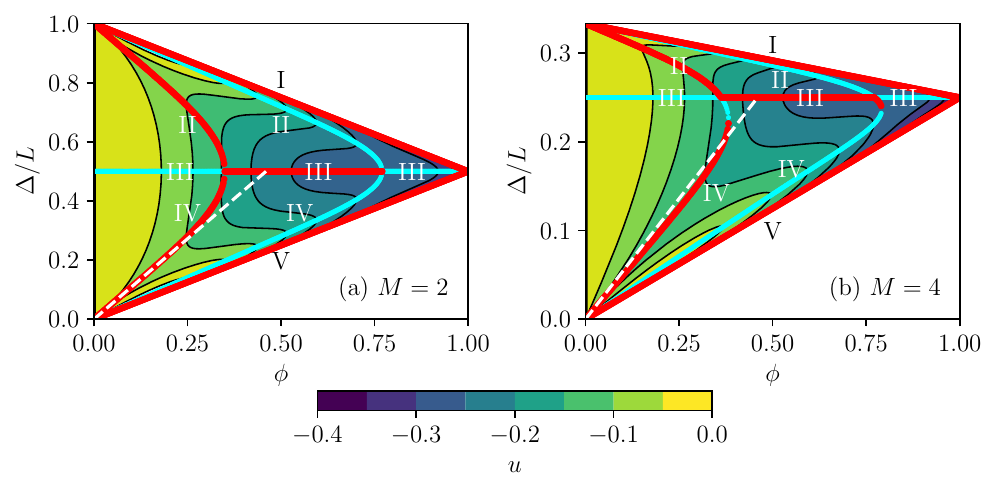}
                \caption{The average energy per particle $u$, given by Eq.~\eqref{eq:Mclustersdelta}, as a function of $\phi$ and $\Delta/L$ for the Siren potential, for (a) $M=2$ and (b) $M=4$. The red and blue curves indicate local minima and maxima, respectively. The white dashed lines indicate the position of the secondary minimum in the Siren potential ($\Delta/D=r_\mathrm{min}/D\approx2.18$), as shown in Fig.~\ref{fig:ab_cb}. The Roman numerals (I-V) correspond to the five distinct configurations illustrated for the case with $M=4$ in Fig.~\ref{fig:states}.}
                \label{fig:BifurcationsSiren}
            \end{figure*}

\begin{figure}[ht]
                \includegraphics[width=\linewidth]{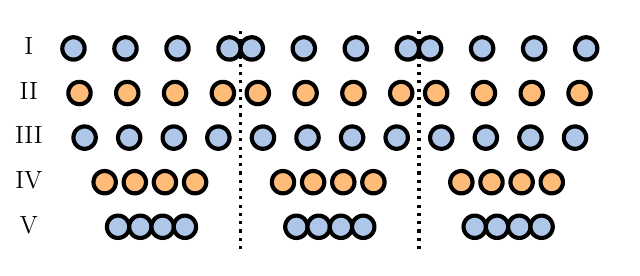}
                \caption{Schematic overview of the five distinct configurations corresponding to the extrema in Fig.~\ref{fig:BifurcationsSiren}, here with $M=4$ and $\phi=0.5$. The dotted lines indicate the edges of the unit cell.}
                \label{fig:states}
            \end{figure}

            We now consider this average energy $u_M$ as a function of $\phi=MD/L$ and $\Delta/L$ for the Siren potential, as shown in Fig.~\ref{fig:BifurcationsSiren}. For two different values of $M$, this figure highlights the (local) minima (in red) and maxima (in blue) on top of the energy landscape. These extreme values correspond to five distinct particle configurations shown in Fig.~\ref{fig:states}.
            The two diagonal lines bounding the wedge (i.e., the limiting values of $\Delta/L$) are determined by $M$, $L$, and $D$. The configurations along these boundaries always minimize the potential energy.
            The maximum value, $\Delta/L=(1-D/L)/(M-1)$, corresponds to the particles at the edges of the unit cell touching those in the neighboring cells (configuration I).
            The minimum value, $\Delta/L=D/L$, corresponds to the clusters of touching particles (configuration V). 
            Moreover, the horizontal lines in Fig.~\ref{fig:BifurcationsSiren} represent $\Delta/L=1/M$, for which the particles are evenly distributed over the unit cell (configuration III). In this case, Eq.~\eqref{eq:Mclustersdelta} is equivalent to the uniform distribution of single particles ($M=1$), given by Eq.~\eqref{eq:u1} and shown in Fig.~\ref{fig:schematic}(a).
            Furthermore, the blue and red curved lines in Fig.~\ref{fig:BifurcationsSiren} correspond to (local) extrema in $u_M$, which are associated with the maxima or minima of the particular potential (configurations II and IV).
    
            For $M=2$, in Fig.~\ref{fig:BifurcationsSiren}(a), the energy landscape is symmetric with respect to the line $\Delta/L=1/M=1/2$. The clustered states along the edges (red lines, configurations I and V) are always stable. 
            At low $\phi$ values, the additional stable equilibria correspond to pairs of particles near the potential's (secondary) minimum. 
            However, for $0.35\lesssim \phi \lesssim 0.77$, the density is sufficiently large, preventing the particles from being at the spacing associated with the minimum of the potential. As a result, the particles repel each other, leading to the uniform distribution ($\Delta/L=1/2$, configuration III) being a stable configuration. 
            The uniform distribution is again unstable at higher densities, leaving the clustered configuration as the only stable option.
            The transitions at $\phi \approx 0.35$ and $0.77$ correspond to a supercritical and subcritical pitchfork bifurcation, respectively\footnote{A supercritical pitchfork bifurcation describes a transition from one stable equilibrium to two stable equilibria (or vice versa). The stability is reversed for the subcritical pitchfork bifurcation \citep{strogatz2018nonlinear}.}.
    
            When $M>2$, in Fig.~\ref{fig:BifurcationsSiren}(b), the symmetry of the wedge is broken. The clustered state (lower diagonal, configuration V) is no longer equivalent to the state with maximum interparticle spacing (upper diagonal, configuration I). An additional asymmetry arises around the bifurcation points. The intersections between the curves (configurations II and IV) and the horizontal line (configuration III) are no longer symmetric with respect to the horizontal line. This asymmetry is a sign of hysteresis upon varying the value of $\phi$: the preparation of the system in terms of $\phi$ and initial particle distribution can lead to different stable states, as further explored in Sec.~\ref{sec:MC1D}.
            For example, starting from a uniform distribution ($\Delta/L=1/M=1/4$) with $\phi\approx0.5$ and gradually decreasing its value (e.g., by increasing the domain size in an experiment), the system follows the upper branch with relatively large particle spacing compared to the lower branch. Similarly, upon gradually increasing $\phi$, the value of $\Delta$ briefly decreases after the bifurcation before it jumps towards the clustered state on the lower edge. In principle, the system should always transition to the lower edge since it avoids crossing a maximum in the energy landscape, while this would be required to go to the upper edge. Such preferential transitions are absent for the symmetric $M=2$ cases.
            
            Moreover, we have indicated the expected equilibrium spacing based on the secondary minimum $r_\mathrm{min}/D\approx2.18$ by the white dashed lines in Fig.~\ref{fig:BifurcationsSiren}. The particles settle at a slightly smaller spacing for small $\phi$ and $M>2$, as indicated by the lower red branch.
            This deviation is attributed to the potential accounting only for the interactions between two particles. However, the energy calculation for this figure includes multiple particles in the same unit cell ($M>2$) over an infinite array of unit cells.

\section{One-dimensional Monte Carlo simulations}\label{sec:MC1D}
    In the previous section, we have gained a basic understanding of the energy landscape near the equilibrium states for systems with only a few particles. There are often multiple local minima for a given particle number density, which additionally show hysteresis in the transitions between equilibrium states. The particle configurations thus depend on the initial conditions and the noise level, which we explore further in this section.
    
    \subsection{Monte Carlo method}
        We use the conventional Metropolis Monte Carlo (MC) algorithm to simulate a system of $N$ particles in a one-dimensional domain with periodic boundaries \citep{frenkel2002understanding}. The domain length $L$ is determined by the user-defined value of the packing fraction $\phi=ND/L$. Each simulation consists of $10^5$ MC steps, where in each step, one particle is randomly selected and moved with a random step size between $-0.1D$ and $0.1D$. The maximum step size is kept constant throughout the simulation.
        
        Before accepting a proposed step, several checks are made to prevent particle overlap, thereby accounting for the hard-sphere repulsion, and to determine the total energy of the new state. The new state is automatically accepted if it has lower energy than the old state and no particle overlap.
        A proposed step resulting in an energy increase $\Delta E$ is accepted with probability
        \begin{equation}\label{eq:MCstep}
            p = e^{-\Delta E/(k_BT)},
        \end{equation}
        where $k_B$ is the Boltzmann constant, and $T$ is the temperature \citep{grimmett2020probability}.
        
        By setting $k_BT>0$, we incorporate (thermal) fluctuations into the system (as commonly included when simulating colloids \citep{almudallal2011simulation}), which govern the amount of noise in the particle motion. 
        It is worth noting that the value of $k_BT$ in our simulations does not reflect the actual thermodynamic temperature of the system but is used to capture the noise also present in the experiments (due to, e.g., particle collisions, fluctuations in the flow, or inhomogeneity of the bottom over which the particles move).
        Furthermore, a positive value of $k_BT$ prevents the system from getting trapped in shallow, local minima of the energy landscape.
    
        Finally, we allow for cluster moves to increase the efficiency of the simulations. Before accepting a move, we define a $20\%$ chance that the neighboring particle to the right will also be moved by the same amount\footnote{The final configuration is not sensitive to the exact percentage; the clusters moves here are primarily used for efficiency, such that clusters are more likely to move.}. This procedure is applied iteratively: if the neighboring particle is added to the move, there is a subsequent $20\%$ chance to add the next neighbor. This loop is truncated once a particle is not added to the cluster move. This procedure satisfies detailed balance, such that the step is reversible with equal probability \citep{grimmett2020probability}. 
    
    \subsection{Equilibrium states and the role of initial conditions}
\begin{figure*}
            \includegraphics[width=0.8\textwidth]{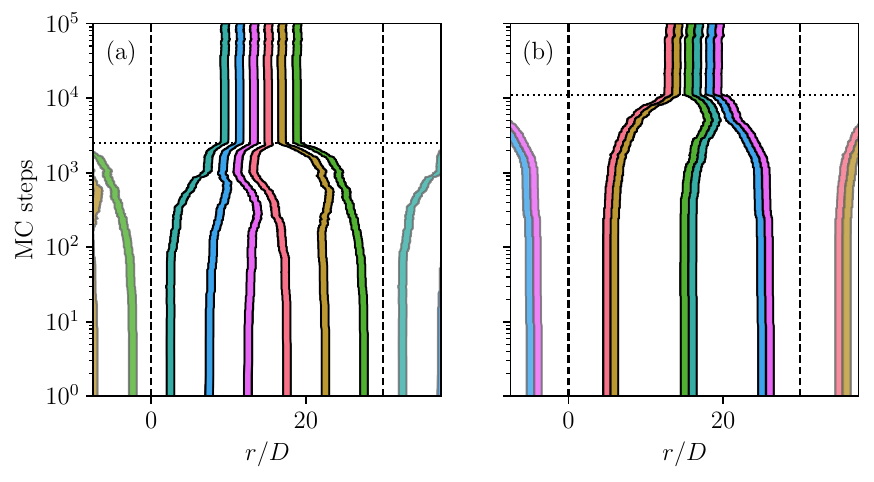}
            \caption{Particle positions over time obtained from 1D Monte Carlo simulations using the Siren potential for $N=6$, $\phi=0.2$, and $k_BT=0$. The simulations are initiated from either a uniform distribution of (a) single particles or (b) pairs of particles. The vertical dashed lines indicate the periodic boundary, and the horizontal dashed line indicates the time at which the system has reached an equilibrium state (after which the particle positions relative to each other remain constant).}
            \label{fig:Siren_1D_examples}
        \end{figure*}

        We begin by examining the equilibrium states of the one-dimensional system governed by the Siren potential without thermal fluctuations ($k_BT=0$). To illustrate the self-organization at the particle level, we show the results of two small simulations, each with $N=6$ particles and $\phi=0.2$, as shown in Fig.~\ref{fig:Siren_1D_examples}. 
        Starting from an initial uniform distribution, in Fig.~\ref{fig:Siren_1D_examples}(a), the particles form structures with a spacing of approximately $r/D\approx1.9$. This value roughly corresponds to the secondary minimum in the potential at $r_\mathrm{min}/D\approx2.2$, but is slightly smaller, as previously predicted based on Fig.~\ref{fig:BifurcationsSiren}. 
        Next, for a uniform distribution of touching pairs, in Fig.~\ref{fig:Siren_1D_examples}(b), the pairs are stable and stay together throughout the simulation.
        Once the system reaches an equilibrium state, the pairs form a compact pattern similar to that of the single particles, with the distance between the centers of the pairs approximately $r/D\approx2.7$. The difference in final states (arrays made of single particles versus pairs) confirms that, at low temperatures, the system is not ergodic.
        
\begin{figure*}
            \centering
            \includegraphics[width=\linewidth]{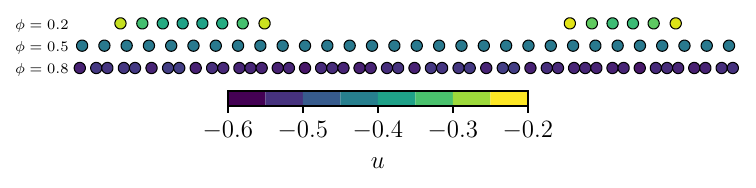}
            \caption{Resulting particle distributions from the 1D MC simulations after $10^5$ steps for $\phi=[0.2,0.5,0.8]$. The colors indicate the total potential energy of each particle in the shown configuration. A comparison of the particle distributions in different potentials (Well, Mermaid, and Siren) is given in Fig.~\ref{fig:Siren_1D_comparison} in the Appendix \ref{sec:appendix}.}
            \label{fig:Siren_1D_overview}
        \end{figure*}

        Next, Fig.~\ref{fig:Siren_1D_overview} gives a comprehensive overview of one-dimensional MC results for three values of $\phi=[0.2,0.5,0.8]$.
        At low $\phi$ values, the particles form patterns with an intrinsic spacing that (approximately) corresponds to the (secondary) minimum in the potential energy (related to configuration IV in Fig.~\ref{fig:states}).
        For larger values of $\phi$, beyond the bifurcation transition in Fig.~\ref{fig:BifurcationsSiren}, the particles arrange themselves equidistantly at relatively small spacing (configuration III).
        Continuing to increase $\phi$ yields the formation of clusters due to short-range attraction (similar to configuration V).
        
        The final states shown in Fig.~\ref{fig:Siren_1D_overview} are remarkably similar to the patterns in the hydrodynamic experiments in Fig.~\ref{fig:example-of-patterns}. 
        The behavior in the green region (with chains and bands) is captured by the Siren potential, where particles maintain an intrinsic spacing until reaching a critical density, after which multiple-particle-wide clusters form. 
        Moreover, the blue region (with chains and irregular clusters) can effectively be described by omitting the short-range attraction (setting $C=0$ in Eq.~\eqref{eq:sirenpotential}) such that multiple-particle-wide clusters are always unstable. See the comparison of particle distributions in different potentials in Fig.~\ref{fig:Siren_1D_comparison} in the Appendix \ref{sec:appendix}.
            
\begin{figure}
            \includegraphics[width=\linewidth]{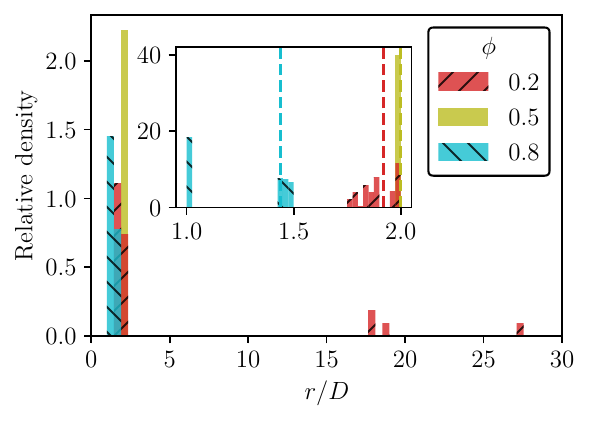}
            \caption{Normalized histogram of the distances between the particles for the Siren potential simulations in Fig.~\ref{fig:Siren_1D_overview}, averaged over the final $10^4$ MC steps, with bin size $\Delta r/D=0.45$. The dashed lines indicate the median distances at $r/D\approx [1.92,2.00,1.44]$ for $\phi=[0.2,0.5,0.8]$, respectively. The insert shows the relative distribution at short distances, with bin size $\Delta r/D=0.025$.}
            \label{fig:hist}
        \end{figure}

        The similarities between the simulations and experiments are further supported by the distributions of particle spacings shown in Fig.~\ref{fig:hist}. At low values of $\phi$ (e.g., $\phi=0.2$), the distances are distributed around the secondary minimum, with a median (normalized) distance of $1.92$. As $\phi$ increases ($\phi=0.5$), the distribution shifts towards a narrow peak, corresponding to a uniform distribution with a median distance of $2.00$. Increasing $\phi$ further (to $\phi=0.8$), we find a bimodal distribution with a clustered and unclustered population with a median distance of $1.44$. In the latter case, some particles form clusters (with $r/D=1$), so the other particles have a larger spacing shifted towards the secondary minimum. 
        
    \subsection{Effect of fluctuations}
        We introduce fluctuations into the system by considering cases with $k_BT>0$. The magnitude of these fluctuations is described by the typical thermal energy $k_BT$ normalized by the maximum value of the potential $U_\mathrm{max}\approx0.096$, as discussed in Sec.~\ref{sec:modelpotential}. Our simulations cover a wide range of parameter values, with $k_BT/U_\mathrm{max}$ spanning from $0.01$ to $2.0$, and coverage fractions $\phi$ ranging from $0.1$ to $1.0$.
        
        At the end of each MC simulation, the distribution of interparticle distances is used to identify different states. Specifically, we define a `densely clustered' state if more than $50\%$ of the distances are smaller than $1.05$. 
        Similarly, a state is classified as `loosely clustered' if more than $50\%$ of the distances fall between $1.8$ and $2.3$. These distances are evenly spaced around the secondary minimum at $r/D\approx2.2$ and close to the intrinsic spacing $r/D\approx1.9$ observed in, e.g., Fig.~\ref{fig:Siren_1D_examples}).
        Lastly, we define an `equidistant' state if more than $50\%$ of the distances lie within the range $[1/\phi-0.1,1/\phi+0.1]$ (since the equidistant spacing is given by $1/\phi$).
        The threshold of $50\%$ defines phases where most particles are within a specific state. Different threshold values shift the `boundaries' of the different phases, especially for larger $k_BT$, but do not qualitatively alter the phase diagram.
    
        \begin{figure}
            \includegraphics[width=\linewidth]{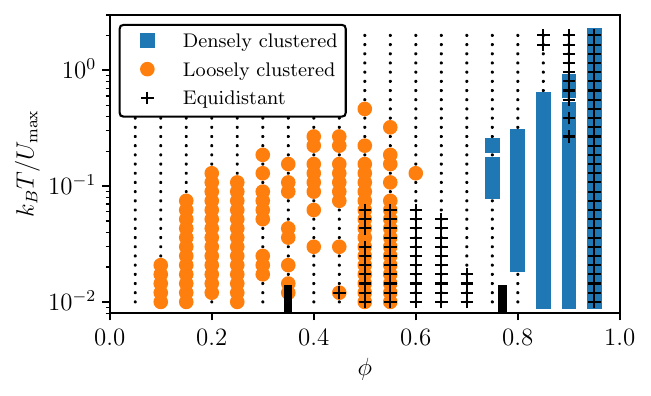}
            \caption{Phase diagram illustrating the different states in the one-dimensional system governed by the Siren potential. The control parameters are the particle coverage fraction $\phi$ and the normalized thermal energy $k_BT/U_\mathrm{max}$. The different states are identified based on the criterion that at least 50\% of the interparticle distances satisfy the corresponding condition. The black dots indicate simulations where no dominant state is found. 
            The black vertical bars at $\phi=0.35$ and $0.77$ correspond to the bifurcation points in Fig.~\ref{fig:BifurcationsSiren}(a) for two particles per unit cell ($M=2$).}
            \label{fig:phasediagram1D}
        \end{figure}

        These states are indicated in the phase diagram in Fig.~\ref{fig:phasediagram1D} as a function of $\phi$ and $k_BT/U_\mathrm{max}$. The phase diagram clearly shows distinct regions corresponding to densely clustered states, loosely clustered states, equidistant states, and states where no configuration is dominant. 
        
        At low values of $k_BT/U_\mathrm{max}$ and $\phi$, a loosely clustered state is found akin to the previous Monte Carlo simulations in Fig.~\ref{fig:Siren_1D_overview}. As $\phi$ increases, the particles are gradually pushed out of the secondary equilibrium towards smaller interparticle distances. After a region where neither state (loosely clustered or equidistant) is dominant, the particles eventually form an equidistant pattern around $\phi\approx0.5-0.7$. Further increasing $\phi$ leads to particles overcoming the potential barrier and forming a densely clustered state. Note that the equidistant state can overlap with the other states since our chosen criteria for the different states are not mutually exclusive. For example, for $\phi=0.50-0.55$, the equidistant spacing $1/\phi=1.8-2.0$ is close to the secondary minimum of the potential. 
        The transitions between the states at low temperatures are accurately predicted by the two black bars at $0.35$ and $0.77$ in Fig.~\ref{fig:phasediagram1D}, corresponding to the bifurcation points illustrated in Fig.~\ref{fig:BifurcationsSiren}(a) for two particles per unit cell ($M=2$) \footnote{Note that $M=2$ does not imply a system with only two particles. Instead, it only implies a maximum cluster size of two.}.
        
        At larger values of $\phi$, densely clustered states are found. Remarkably, at $\phi=0.8$, the clustered states only form within a specific temperature range: $0.03\lesssim k_BT/U_\mathrm{max}\lesssim0.2$. Below this range, there are still some clusters and some particles at large distances, i.e. hysteresis is relatively important here. However, the thermal energy is insufficient to enable more particles to overcome the potential barrier and cluster together. 
        Conversely, for sufficiently large values of $k_BT/U_\mathrm{max}$, the variations in the potential are only minor compared to the thermal energy. As a result, the particles do not strongly feel the potential, and their dynamics are not significantly influenced. The interparticle distances in these cases are more randomly distributed, analogous to a gas-like phase in higher dimensions.
    
        Concerning the hydrodynamic experiments, we can capture the behavior in the blue and green regions of the parameter space in Fig.~\ref{fig:example-of-patterns} using the Siren potential in a one-dimensional system. Nonetheless, the orange-shaded region (characterized by isolated particles and mostly two-particle-wide bands) is not effectively captured by such one-dimensional modeling, as the structures in this region are two-dimensional. The following section addresses such two-dimensional cases and further relates our efforts to similar colloidal systems.

\section{From individual particles to patterns}\label{sec:MC2D}
    So far, we have focused on the one-dimensional case governed by the Siren potential, representing the interactions between already-existing parallel chains in the hydrodynamic system.
    We now use a two-dimensional approach to simulate individual particles instead of entire chains to obtain a more complete representation of the physical system. This extension allows for including the formation of the chains and bands in the simulations, i.e., the stage before the system can be described as one-dimensional.
    Furthermore, this approach allows us to address which patterns from the hydrodynamic system can also be replicated in a colloidal system.
    
    \subsection{Two-dimensional model potential}
        We propose a two-dimensional model potential inspired by the dipolar structure of the steady streaming flow around a single particle and additionally based on polarizable colloidal particles at a fluid-fluid interface.
        The deformation of the interface around each colloid induces capillary interactions between them. Following the description by \citet{kralchevsky2000capillary}, the capillary force between two particles at an interface is inversely proportional to their separation $r$, for interparticle distances which are large compared to the particle size but small compared to the capillary length (approximately \SI{2.7}{mm} for an air-water interface). Both criteria are satisfied for long-range attraction between colloidal particles at an air-water interface.
        An external oscillating electric field parallel to the interface induces a dipole moment in each particle, with the resulting interactions also parallel to the interface.
        The potential of each particle is then described using the dipole-capillary (dc) potential
        \begin{equation}\label{eq:dipolecapillary}
        \begin{split}
            U_{dc}(r/D,\theta) &= U_\mathrm{dipolar} + U_\mathrm{capillary} \\
            &= \frac{A_d}{\left(r/D\right)^3}\left(1-3\cos^2\theta\right) + A_c\ln{\left(r/D\right)},
        \end{split}
        \end{equation}
        where $A_d$ is a positive constant controlling the formation of chains and bands and their interactions at close range (similar to \citep{almudallal2011simulation}), $A_c$ is a positive constant that sets the strength of the long-range attractions due to capillary effects, and $\theta$ is the angle with respect to the direction of the electric field. Note that $U_{dc}$ does not converge to a constant value as $r/D\rightarrow\infty$. While this may not be strictly physical, it serves nevertheless as a useful model for our analysis. The corresponding force scales as $1/r$ at long ranges and thus goes to zero at infinity.
        We further assume that the interactions are pairwise additive and that the induced dipoles do not induce higher-order effects. 
    
        The characteristics of the potential in Eq.~\eqref{eq:dipolecapillary} are governed by two control parameters, but the specific shape is determined only by the ratio $A_c/A_d$. Such a single control parameter that sets the shape is analogous to the potential observed in the hydrodynamic system. There, the shape of the potential depends on $A_r/D$, which determines the strength and positions of the vortices in the time-averaged flow (shaping the hydrodynamic forces on the particles).
    
\begin{figure}
            \includegraphics[width=\linewidth]{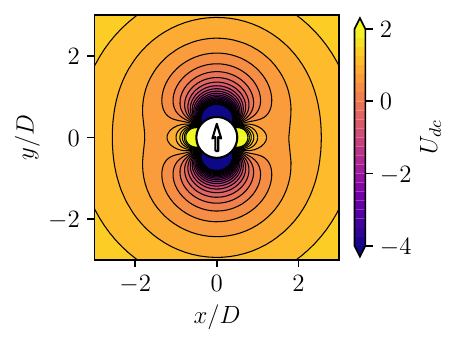}
            \caption{(Color online) Example of the dipole-capillary potential described by Eq.~\eqref{eq:dipolecapillary}, incorporating dipolar and capillary interactions, with $A_c=A_d=1$. The colors correspond to the values of $U_{dc}$ and are separated by black contour lines with steps of $0.25$. The arrow represents the direction of the dipole moment.}
            \label{fig:2DExample}
        \end{figure}

        Figure~\ref{fig:2DExample} shows an example of the potential described by Eq.~\eqref{eq:dipolecapillary} around a single particle with $A_c=A_d=1$. Note that the potential is strongly attractive along the direction of the dipole moment, i.e., in the $y$-direction. Conversely, along the $x$-axis, the potential has a (local) minimum where the dipole repulsion and capillary attraction balance each other.
    
\begin{figure*}
            \includegraphics[width=\linewidth]{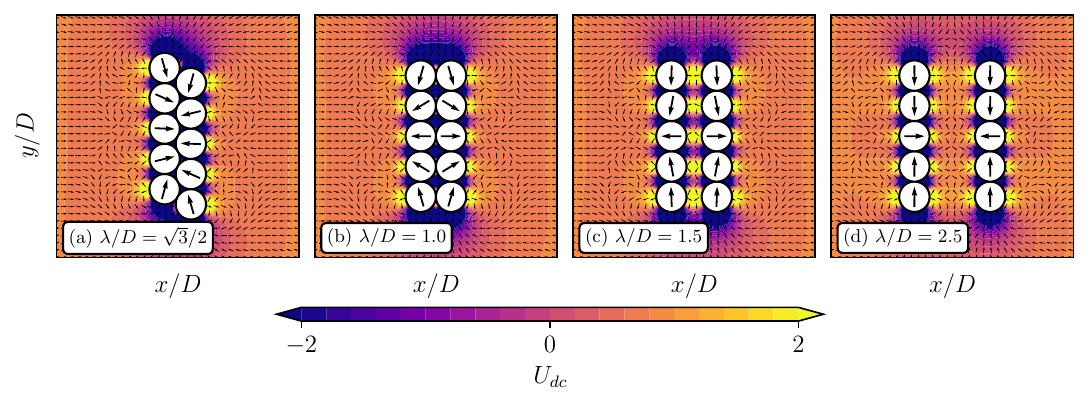}
            \caption{(Color online) Examples of the interaction between two parallel particle chains, each consisting of five particles, at various normalized spacings $\lambda/D$, for $A_d=1$ and $A_c=0.05$. The colors represent the values of the potential given by Eq.~\eqref{eq:dipolecapillary}. The black arrows indicate the direction of the forces, with the larger arrows being the forces at the centers of the particles.}
            \label{fig:2D2chains}
        \end{figure*}

        In Fig.~\ref{fig:2D2chains}, we consider a configuration with two parallel particle chains with five particles each. We set $A_d=1$ and $A_c=0.05$, so the capillary attraction dominates only at long ranges. Overall, the interactions resemble those previously described for both the Siren potential and the hydrodynamic experiments. 
        For the staggered configuration in Fig.~\ref{fig:2D2chains}(a), where the chains are shifted by $D/2$ relative to each other in the $y$-direction, the chains experience a relatively strong attractive force, as previously observed for dipolar interactions in this configuration \citep{almudallal2011simulation}. 
        Conversely, for an aligned configuration in Fig.~\ref{fig:2D2chains}(b), the force between the chains is strongly repulsive instead. Also, for small chain spacings $\lambda/D$, in Fig.~\ref{fig:2D2chains}(c), the interaction is repulsive at mid-ranges (due to dipolar repulsion). Increasing $\lambda/D$ further in Fig.~\ref{fig:2D2chains}(d) leads to a net attraction between chains at long ranges due to capillary effects.
    
\begin{figure}
            \includegraphics[width=\linewidth]{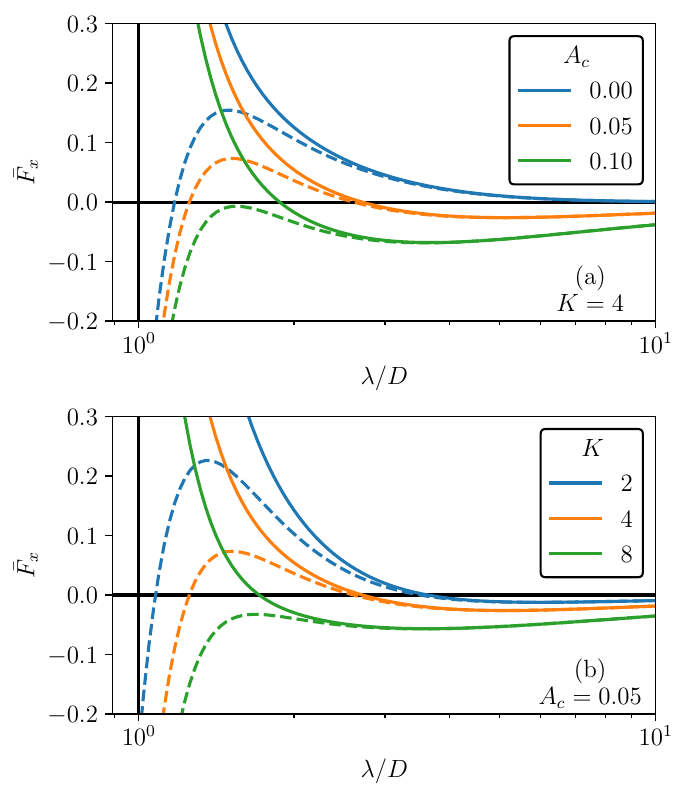}
            \caption{The average of the force $F_x$ on each particle (in the right chain) for a configuration with two parallel chains in straight (solid) and staggered (dashed) configurations, with $A_d=1$. (a) Variation of $A_c$ while keeping $K=4$ fixed. (b) Variation of $K$ while keeping $A_c=0.05$ fixed. Positive values of $\bar{F}_x$ correspond to repulsion between the chains.}
            \label{fig:2DPotential_LA}
        \end{figure}

        The interaction between two parallel chains depends on both $A_c/A_d$ and the specific arrangement of the chains. To gain a better understanding, we show the $x$-component of the force averaged over all particles within a chain, $\bar{F}_x$, in Fig.~\ref{fig:2DPotential_LA}. We vary the values of $A_c$ (with $A_d=1$ fixed), the normalized chain spacing $\lambda/D$, the number of particles $K$ within each chain, and the configuration (staggered or straight). 
    
        For $A_c>0$ and a fixed value of $K$, such as $K=4$ in Fig.~\ref{fig:2DPotential_LA}(a), the interaction is always attractive at sufficiently large spacings. Note that for $A_c=0$, long-range attraction is absent. Moreover, at these large spacings, the difference in the average force between the straight and staggered configurations becomes negligible. However, as the spacing decreases, the configuration significantly impacts the interaction. At very short ranges, with (almost) touching chains, the force is either strongly repulsive for straight chains or attractive for staggered chains. Furthermore, for certain combinations of $K$ and $A_c$, e.g., for $K=4$ and $A_c=0.05$, there is an intermediate repulsive range for staggered chains.
    
        Remarkably, for a fixed value of $A_c$, e.g., for $A_c=0.05$ in Fig.~\ref{fig:2DPotential_LA}(b), increasing the chain length $K$ has a similar effect on the average force as increasing the capillary attraction. This can be attributed to the increasing number of interactions between particles in neighboring chains as $K$ increases. 
        For long chains, the value of $\theta$ (the angle between the line connecting two particle centers and the electric field direction) between two particles in neighboring chains is typically close to $0$ or $\pi$, such that the force is attractive. 
        Contrarily, for shorter chains, values of $\theta$ closer to $\pi/2$ are relatively more important, leading to a strongly repulsive force.
    
        The results in Fig.~\ref{fig:2DPotential_LA}(b) imply that the magnitude of the long-range attraction increases with chain length and thus increases over time during the self-organization of single particles into chains. These results also show that the chain spacing $\lambda$, closely related to the zero crossings of the solid lines in Fig.~\ref{fig:2DPotential_LA}, depends on $K$ and thus changes dynamically over time, even if $A_d$ and $A_c$ are kept constant.
        In contrast, the chain length does not influence the equilibrium spacing for the hydrodynamic system, especially not for longer chains. The influence of the chain length on the system's interactions is thus a fundamental difference between the hydrodynamic and these model systems.
    
        In summary, Fig.~\ref{fig:2DPotential_LA} demonstrates that the potential described by Eq.~\eqref{eq:dipolecapillary} can produce interactions between chains that are similar to those governed by the Siren potential. However, these specific results in Fig.~\ref{fig:2DPotential_LA} only hold for particles that are already arranged in parallel chains. Next, we briefly explore the formation and subsequent dynamics of the chains in such 2D systems (including the sensitivity to initial conditions), again resorting to Monte Carlo simulations.
    
    \subsection{Two-dimensional Monte Carlo simulations}
        Now, we consider Monte Carlo simulations of a two-dimensional system governed by the dipole-capillary potential to replicate some of the typical patterns observed in the hydrodynamic experiments, as shown in Fig.~\ref{fig:example-of-patterns}. We use a code similar to that used in Sec.~\ref{sec:MC1D}.
        We use a domain of size $(L_x\times L_y) = (20\times10)D$ with a periodic boundary in the $x$-direction and an impenetrable boundary in the $y$-direction. This choice intentionally limits the maximum chain length to ten particles and allows us to adjust $A_c$ accordingly to obtain a Siren-like interaction between parallel chains (recall that longer chains induce stronger attractive interactions, as depicted in Fig.~\ref{fig:2DPotential_LA}).    
    
\begin{figure*}
            \includegraphics[width=\textwidth]{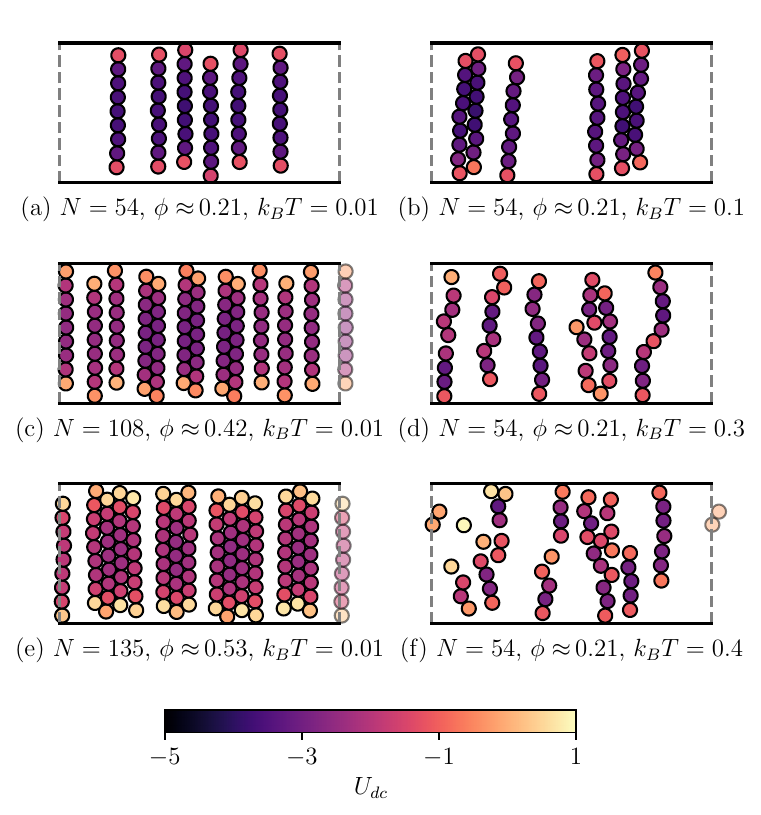}
            \caption{Particle positions after $10^5$ MC steps, starting from parallel chains, with $A_d=1$ and $A_c=0.01$ both constant. We either increase $N$ at constant $k_BT$ (left column) or increase $k_BT$ at constant $N$ (right column). The colors indicate the energy of each particle within the potential landscape.}
            \label{fig:MC2D_NT}
        \end{figure*}

        In Fig.~\ref{fig:MC2D_NT}, we specifically fix the values of $A_d=1$, $A_c=0.01$, while varying the number of particles in the domain. At the start of each simulation, the particles are already in parallel chains to prevent clustering during chain formation\footnote{Preliminary simulations show that starting from a random initial condition leads to enhanced clustering.}.
        To prevent the system from getting trapped in shallow local minima, we set $k_BT=0.01$, i.e., small with respect to both $A_d$ and the local extrema of the potential [see, e.g., Fig.~\ref{fig:2DExample}]. 
        At low values of $\phi$, as shown in Fig.~\ref{fig:MC2D_NT}(a), the chains maintain their preferential spacing without spreading out. As we gradually increase $\phi$, depicted in Figs.~\ref{fig:MC2D_NT}(c)~and~(e), the chains form wider bands, resembling the behavior observed in systems governed by a Siren potential. Specifically, this behavior corresponds to the green-shaded region in Fig.~\ref{fig:example-of-patterns}.
    
        Using the dipole-capillary potential, we cannot reproduce the result from the blue-shaded region in Fig.~\ref{fig:example-of-patterns}, characterized by predominantly one-particle-wide chains and some irregular clusters, without any multiple-particle-wide bands. As we already learned in the previous sections, this pattern only occurs if there is no short-range attraction, with the interactions described by the Well potential\footnote{We additionally recall from \citet{overveld2023pattern} that for $A_r\lesssim0.7$, the inner vortices in the time-averaged flow are small, leading to no short-range attraction and the two-particle-wide bands being unstable.}. However, the short-range attraction in the dipole-capillary potential is inherently linked to the dipole interaction and is thus always present. Reproducing the hydrodynamic patterns in the blue-shaded region is thus not feasible with the specific dipolar interactions used in our 2D model system.
    
        Contrarily, we can attempt to reproduce the pattern in the orange-shaded region in Fig.~\ref{fig:example-of-patterns}, characterized by isolated particles and mostly two-particle-wide bands. The instantaneous interactions between particles are weak in this part of the parameter space, relative to the noise due to, e.g., small inhomogeneities in the flow field or in the bottom over which the particles move. The weak interactions destabilize particle pairs [see \citet{overveld2022numerical}] and likely also particle chains [see \citet{overveld2023pattern}].
        We try to reproduce this more disordered state by increasing the influence of thermal fluctuations on the system, as shown in the right column of Fig.~\ref{fig:MC2D_NT}.
        The increase in $k_BT$ in Figs.~\ref{fig:MC2D_NT}(b)~and~(d) allows some particles to overcome the potential barrier and form bands. Moreover, the fluctuations lead to tortuosity in the chains and introduce some defects.
        As we raise the temperature further, i.e., in Fig.~\ref{fig:MC2D_NT}(f), the chains break, resulting in a mixture of dispersed single particles and more coherent clusters. This state resembles the hydrodynamic experiments conducted at $A_r/D\approx2$ and $\phi\approx0.17$, i.e., the orange-shaded region in Fig.~\ref{fig:example-of-patterns}.
    
\section{Discussion}\label{sec:discussion}
    Most characteristic patterns of the hydrodynamic experiments from \citet{overveld2023pattern} can be captured in the 2D MC simulations using the model potential [Eq.~\eqref{eq:dipolecapillary}].
    However, we cannot replicate all the different distinct patterns observed in the experiments, strongly suggesting that the hydrodynamic interactions have additional complexity. 
    In particular, the dipolar-capillary model is always attractive at short distances (for the staggered configuration), limiting the range of different cases that can be captured. Removing (or negating) such a feature requires adding more complexity to the model. In contrast, the vortex-induced interactions in the hydrodynamic system already include this complexity: i.e., the interactions at different ranges (short, mid, long) are all intrinsically present. Another crucial aspect is the difference between the nonlinear interactions in the hydrodynamic system versus the linear interactions in our 2D potential based on a colloidal system. Moreover, in our simplified model, the inherently multibody interactions are simplified using the assumption of pairwise additive interactions.
    
    % Note about nonlinearity
    The interactions in the hydrodynamic system are induced by the vortices in the steady streaming flow. Their size, strength, and position are predominantly influenced by the system's geometry, as it is specifically related to the wake shed by the objects in the flow \citep{overveld2023pattern}. When two chains merge into a two-particle-wide band, the time-averaged flow is not merely the superposition of the two flow fields around the single chains. Consequently, the net interactions also combine in a nonlinear way, such that the net interactions with a two-particle-wide band are not a simple summation of the interactions with two separate chains.
    
    In contrast, in our 2D MC simulations, we assume that the net interactions are due to a linear superposition of potentials from each individual particle. As a result, forces associated with two-particle-wide bands are approximately twice as large compared to single chains. During the formation of a pattern, this leads to self-enhanced clustering as the long-ranged attractive forces increase as the structures grow. 
    These differences emphasize that nonlinear interactions, even for similar potentials of the single particle, can fundamentally alter self-organization processes.  
    
    % Tuning interactions and control
    The linear interactions additionally have implications for the control of the patterns. 
    The dipole-capillary potential in the 2D MC simulations requires careful tuning of the value of $A_c$ in relation to the chain length $K$ since the interactions between two aligned chains depend on the number of particles in each chain. Moreover, confining the system along the direction of the chains is essential to prevent excessively long chains from forming, as their interactions would always become attractive when they grow sufficiently long. This confinement also restricts the mobility in the direction of the chains, resulting in predominantly one-dimensional dynamics. 
    For future work, rather than being a limitation, this confinement constraint can be advantageous in an experiment where the domain boundaries are controlled. The manipulation of the boundaries directly leads to active control over the strength of the interactions within the system. Such an approach of manipulating self-organization through confinement is gaining significant research interest \citep{araujo2023steering}.

\section{Conclusions}\label{sec:conclusion}
    % Takehome 1D
    In this study, we explored self-organization phenomena in a hydrodynamic system, using tools commonly used to describe colloidal systems. Specifically, we introduced the \emph{Siren} potential - a simplified model potential combining short-range attraction, mid-range repulsion, and long-range attraction, which serves as a generalization of other common potentials like the Mermaid (SALR) or Well potentials. 
    By applying the Siren potential to describe the interactions between parallel particle chains, our Monte Carlo simulations successfully replicate many of the characteristic patterns observed in hydrodynamic experiments. Moreover, this approach yielded a comprehensive phase diagram, providing valuable insight into the system's equilibrium states.
    
    % Takehome 2D
    We expanded our analysis to two-dimensional systems of colloidal particles at an interface, with dipole moments induced by an external oscillating electric field parallel to the interface. We introduced a model potential that incorporates both the dipolar and capillary interactions. 
    Starting from a distribution of parallel chains, the simulations within the two-dimensional colloidal system resemble the patterns observed in several of the hydrodynamic experiments. 
    
    While the model potentials offer promising results, our MC simulations also highlight challenges in designing colloidal experiments with Siren-like interactions. A key factor is confinement, which limits the maximum chain length and, consequently, the net interactions between the chains. While this mechanism typically adds complexity to the system, it also presents opportunities for control. 
    Additionally, we identified that the interactions of separate particles forming chains differ from those between parallel chains. From this, we derive that the nonlinear interactions (such as multiparticle effects) in the hydrodynamic system are important for the distinct steps in the self-organization process. 
    An equivalent nonlinear mechanism is currently not included in the dipole-capillary potential used. However, such an addition (e.g., by changing the interactions based on the cluster size) may be crucial to obtaining the different stages of the pattern formation within a single simulation.
    
    In conclusion, the Siren and dipole-capillary potentials are powerful tools for exploring and understanding self-organization phenomena in various systems. They clarify the dynamic particle behavior due to the complex interactions encountered in both hydrodynamic and colloidal systems. Our research can serve as a proof-of-principle for future studies, providing a foundation for theoretical model systems and their experimental counterparts.

\section*{Data Availability Statement}
The data that support the findings of this study are (soon) openly available in 4TU.ResearchData at https://doi.org/10.4121/4603f510-007a-4354-b40c-8ab4eaca07cf.

\appendix*
\section{Model potential characteristics}\label{sec:appendix}
    We examine the potential $U$ with three contributions, given by 
    \begin{equation}\label{eq:APP_sirenpotential}
        U(r) = -\frac{A}{r^2}+\frac{B}{r^4}-\frac{C}{r^6},
    \end{equation}
    where $r$ is the distance between two particles (here normalized by the diameter), and $A$, $B$, and $C$ are positive constants, representing the magnitude of long-range attraction, mid-range repulsion, and short-range attraction, respectively. Additionally, we define $r=1$ as the smallest possible particle spacing, representing hard-sphere repulsion.
    
    The extreme values of this potential are located at
    \begin{eqnarray}\label{eq:APP_r}
        r_1 = \sqrt{\frac{B-\sqrt{B^2-3AC}}{A}},\\
        r_2 = \sqrt{\frac{B+\sqrt{B^2-3AC}}{A}},
    \end{eqnarray}
    which additionally yields
    \begin{eqnarray}\label{eq:APP_U}
        U_1 = U(r_1) = \frac{2C}{r_1^6}-\frac{B}{r_1^4},\\
        U_2 = U(r_2) = \frac{2C}{r_2^6}-\frac{B}{r_2^4}.
    \end{eqnarray}
    Based on the positions of these extrema and the values of $A$, $B$, and $C$, we can categorize the potential into qualitatively distinct forms used in Sec.~\ref{sec:modelpotential}.
    
    Based on Eq.~\eqref{eq:APP_r}, in order for the values of $r_1$ and $r_2$ to be real, it is necessary that $B^2 - 3AC \geq 0$, which implies $AC \leq B^2/3$. Above this criterion, the long-range attraction dominates over the other interactions such that the potential is attractive at any range.
    
    There is no short-range attraction if the maximum of the potential occurs at a shorter range than the hard-sphere repulsion, i.e., for $r_1<1$. This condition yields $3C\leq(2B-A)$ to obtain a \textbf{Well potential}.
    For $A>2B$, this condition can not hold regardless of the value of $C$ (since all constants are positive), and long-range attraction overcomes the repulsion at all ranges.
    
    Furthermore, there is no long-range attraction if $A=0$. However, for any positive value of $A$, long-range attraction becomes dominant at (very) long distances. Note that, additionally, for a \textbf{Mermaid potential}, short-range attraction should be included, yielding $3C\geq 2B$ for $r_1>1$. 
    
    In all other cases, all three interactions are relevant. The combination of hard-sphere repulsion and $r_1>1$ implies that $r=1$ is a (primary) minimum of the potential. This primary minimum is the global minimum when $3C>2\sqrt{A^2-AB+B^2}-2A+B$, resulting in a potential similar to those used in DLVO theory. 
    Finally, if the secondary minimum is the global minimum, i.e., for $3C\leq2\sqrt{A^2-AB+B^2}-2A+B$, we obtain a \textbf{Siren potential}.

\begin{figure*}
        \centering
        \includegraphics[width=\linewidth]{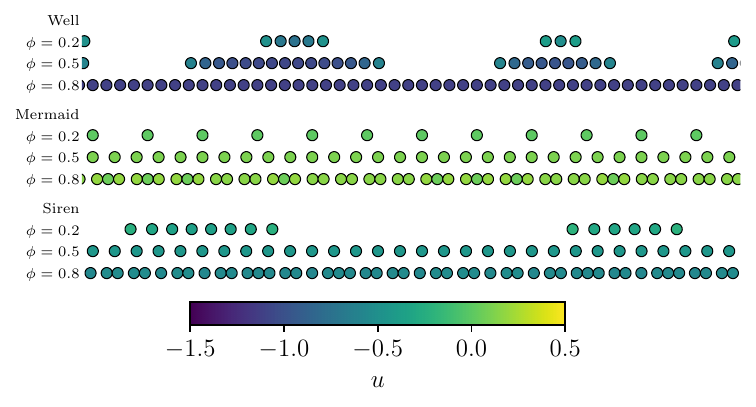}
        \caption{The particle distributions from the 1D MC simulations for the Well, Mermaid, and Siren potentials, as given in Fig.~\ref{fig:ab_cb}. Results are shown after $10^5$ steps for $\phi = [0.2, 0.5, 0.8]$. The colors indicate the total potential energy of each particle in the shown configuration.}
        \label{fig:Siren_1D_comparison}
    \end{figure*}

    For the three aforementioned model potentials, we present in Fig.~\ref{fig:Siren_1D_comparison} the particle distributions obtained from one-dimensional Monte Carlo simulations, as detailed in Sec.~\ref{sec:MC1D}. The values of $A$, $B$, and $C$ are identical to those in Fig.~\ref{fig:ab_cb}(b). 
    Note the similarity between the distribution in the Siren potential and that in the Well potential (with an intrinsically set spacing) or the Mermaid potential (with small clusters of touching particles), for low or high $\phi$ values, respectively.

%merlin.mbs apsrev4-1.bst 2010-07-25 4.21a (PWD, AO, DPC) hacked
%Control: key (0)
%Control: author (8) initials jnrlst
%Control: editor formatted (1) identically to author
%Control: production of article title (-1) disabled
%Control: page (0) single
%Control: year (1) truncated
%Control: production of eprint (0) enabled
%

% \nocite{*}
% \bibliographystyle{unsrtnat}
% \bibliography{main}% Produces the bibliography via BibTeX.

\end{document}